\documentclass[letterpaper]{article} %
\usepackage[]{aaai25} %
\usepackage{times}  %
\usepackage{helvet}  %
\usepackage{courier}  %
\usepackage[hyphens]{url}  %
\usepackage{graphicx} %
\usepackage{natbib}  %
\usepackage{caption} %
\usepackage{algorithm}
\usepackage{algorithmic}
\usepackage{float}
\newfloat{algorithm}{t}{lop}

\usepackage{newfloat}
\usepackage{listings}
\DeclareCaptionStyle{ruled}{labelfont=normalfont,labelsep=colon,strut=off} %
\floatstyle{ruled}
\newfloat{listing}{tb}{lst}{}
\floatname{listing}{Listing}
\usepackage{iftex}

\ifpdf
  \usepackage{underscore}         %
  \usepackage[T1]{fontenc}        %
\else
  \usepackage{breakurl}           %
\fi

\usepackage[utf8]{inputenc} %
\usepackage{url}            %
\usepackage{booktabs}       %
\usepackage{amsfonts}       %
\usepackage{nicefrac}       %
\usepackage{microtype}      %
\usepackage{xcolor}         %
\usepackage{mathtools}

\usepackage{amsmath}

\usepackage{array}
\usepackage{makecell}

\usepackage{comment}

\usepackage{caption} 
\title{Quantifying Misalignment Between Agents: Towards a Sociotechnical Understanding of Alignment}
\author {
    Aidan Kierans\textsuperscript{\rm 1},
    Avijit Ghosh\textsuperscript{\rm 2,1},
    Hananel Hazan\textsuperscript{\rm 3},
    Shiri Dori-Hacohen\textsuperscript{\rm 1}
}
\affiliations {
    \textsuperscript{\rm 1}University of Connecticut\\
    \textsuperscript{\rm 2}Hugging Face\\
    \textsuperscript{\rm 3}Tufts University\\
    aidan.kierans@uconn.edu, avijit.g@uconn.edu, hananel@hazan.org.il, shiridh@uconn.edu
}

\begin{document}

\nocopyright

\maketitle

\begin{abstract}
Existing work on the alignment problem has focused mainly on (1) qualitative descriptions of the alignment problem; (2) attempting to align AI actions with human interests by focusing on value specification and learning; and/or (3) focusing on a single agent or on humanity as a monolith. Recent sociotechnical approaches highlight the need to understand complex misalignment among multiple human and AI agents. We address this gap by adapting a computational social science model of human contention to the alignment problem. Our model quantifies misalignment in large, diverse agent groups with potentially conflicting goals across various problem areas. Misalignment scores in our framework depend on the observed agent population, the domain in question, and conflict between agents' weighted preferences. Through simulations, we demonstrate how our model captures intuitive aspects of misalignment across different scenarios. We then apply our model to two case studies, including an autonomous vehicle setting, showcasing its practical utility. Our approach offers enhanced explanatory power for complex sociotechnical environments and could inform the design of more aligned AI systems in real-world applications.

\end{abstract}

\begin{links}
  \link{Code}{https://github.com/RIET-lab/quantifying-misalignment}
\end{links}

\section{Introduction}
In recent years, growing concerns have emerged about the AI alignment problem~\cite[e.g.][]{russell2019human,christian2020alignment}. With respect to the current AI systems that exist today, Russell~\cite{russell2019human} made the bold yet compelling claim that social media AI today is already misaligned with humanity (e.g. through extensive disinformation spread). As agentic AI systems are growing in importance and increasingly deployed~\cite{kapoor2024aiagentsmatter}, concerns about misalignment of agent-based systems have grown~\cite{chan2023harms,gabriel2024ethicsadvancedaiassistants}. Yet with few exceptions \cite[e.g.][]{critch2020ai}, much prior work focuses on single AI settings \cite{ji2024ai}, on value specification and learning~\cite{critch2020ai,leike2018scalable}, or centers technosolutionist fixes. A healthier alternative would train a sociotechnical lens~\cite{lazar2023ai} on multi-agent settings~\cite{gabriel2024ethicsadvancedaiassistants}, which may include a complex mix of AI agents and humans, potentially holding a dizzying array of competing and often conflicting goals. \citeauthor{critch2020ai} posed the question, ``where could one draw the threshold between `not very well aligned' and `misaligned'[..]?''~\citeyearpar[pg. 14]{critch2020ai}, while \citeauthor{lazar2023ai} call for a ``sociotechnical approach to AI safety,'' stating unequivocally that ``no group of experts (especially not technologists alone) should unilaterally decide what risks count, what harms matter, and to which values safe AI should be aligned'' \citeyearpar{lazar2023ai}. In order to consider the alignment of AI in its sociotechnical environment, we need to expand our definitions of alignment toward a sociotechnical, agent-based understanding that accounts for the complexity of real-world systems in an increasingly AI-agentic world. The question emerges: with whom are agents aligned or misaligned, and on what areas? The alignment of an AI agent or system might be with an adversary, rather than with the developers and/or owners of it; and the alignment---or lack thereof---might be context-, user-, or agent-dependent.

To that end, we propose a novel probabilistic model of misalignment that is predicated on the population of agents being observed (whether human, AI, or any combination thereof), as well as the problem area at hand and, by extension, the agents' values and sense of importance regarding that area. To do so, we extend and adapt a model of contention from computational social science~\cite{jang2017modeling} and apply the adapted model to the alignment problem. 

\textbf{Our contributions} in this paper are as follows: 
\begin{enumerate}
    \item We introduce a novel mathematical model of misalignment which accounts for a population of human and AI agents, and which is parameterized by problem areas, allowing any pair of agents to be simultaneously aligned \textit{and} misaligned. Our model affords explanatory power for the complexity of real-world, sociotechnical settings in which AI agents are deployed.
    \item We simulate a set of worlds with a variety of agents, problem areas, and goals in order to study key drivers of misalignment under our model.
    \item We showcase important features of our model in two case studies that are difficult to account for in previous models, namely, (a) a shopping recommender system and (b) a pre-crash autonomous vehicle (AV) decision-making setting.
\end{enumerate}

The remainder of the paper is organized as follows: after sharing related work (\S{2}), we define the model mathematically (\S{3}). We then describe the setup of our simulation (\S{4}) and discuss the results of different initial settings (\S{5}). Next, we turn to our two case studies (\S{6}), before sharing a discussion (\S{7}) and concluding remarks (\S{8}).

\section{Related Work}
The deployment of powerful agentic AI systems has led to increased concerns about misalignment~\cite{chan2023harms,gabriel2024ethicsadvancedaiassistants}. Several researchers argue that existing AI systems already exacerbate threats to information ecosystems and collective decision-making~\cite{bucknall2022current,dorihacohen2021restoring, russell2019human,Colemane2025764118,SegerEpistemicSecurity}. However, most prior work defines misalignment as a global characteristic of an AI system, often as a binary~\cite{sierravalue2021}. Recent work in machine ethics attempts to answer questions like ``whom should AI align with?'', while others measure alignment with respect to ``human values''~\cite{ji2024ai}. However, while some researchers have noted alignment issues posed by value pluralism~\cite{gabrielartificial2020}, methods for measuring (mis)alignment between different cultures and AI agents remain underdeveloped.

To expand our understanding of misalignment to address this gap, our work draws significant inspiration from a computational model of contention in human populations proposed by \citet{jang2017modeling}. Their model quantifies the proportion of people in disagreement on stances regarding a topic, parameterized by the observed group of individuals. We extend this approach in several key ways: (1) we adapt the model to capture misalignment with respect to goals, rather than stances on topics; (2) we extend the population to include both human and AI agents, allowing for analysis of mixed-agent scenarios; (3) we introduce the concept of ``problem areas'' to segment and analyze alignment across different domains of interaction; (4) we allow for varying levels of goal conflict and importance, providing a nuanced representation of alignment dynamics. While the contention paper focused on controversy in public discourse \cite{jang2017modeling}, our model provides a framework for quantifying misalignment in complex sociotechnical systems where humans and AI agents interact. This approach offers greater real-world explanatory power, with applications in both AI research and policy (see \S{C}).

\section{Modeling misalignment in populations} 

Our approach to modeling misalignment rests on a key observation: understanding and "solving" the AI alignment problem requires first grappling with the challenges of aligning humans. The ubiquity of human conflicts highlights the difficulty of alignment even without AI. Adapting the computational contention model's individualized framing~\cite{jang2017modeling}, we quantify misalignment based on each human and AI agent's goals in problem areas.

\textbf{3.1 \quad Definitions and Notation.}
We define $\mathcal{A} = \{a_1 .. a_n\}$ as a set of $n$ agents and $\mathcal{P} = \{p_1 .. p_m\}$ as a set of $m$ problem areas of interest to at least one agent in $\mathcal{A}$. 
$h(a,g,j)$ is a relationship denoting that agent $a$ holds goal $g$ in problem area $j$. $g^i_j$ is the goal held by agent $i$ in problem area $j$, and $w^i_j$ is the weight, representing the importance that agent $i$ assigns to problem area $j$.

We define $\hat{\mathcal{G}_j} = \{g_1, g_2, .. g_k\}$ as the set of $k$ non-zero goals with regards to problem area $j$ in the set of agents $\mathcal{A}$. We use $g_0$ to denote that an agent holds no goal with respect to problem area $j$:

\begin{equation}
h(a,g_0,j) \iff \not\exists g \in \hat{\mathcal{G}_j} \text{ s.t. } h(a,g,j).
\end{equation}

Let $\mathcal{G}_j = \{g_0\} \cup \hat{\mathcal{G}_j}$ be the set of $k + 1$ extant goals with regard to $j$ in $\mathcal{A}$.

\textbf{3.2 \quad Measuring Conflict.} In order to capture the notion that goals can be compatible or conflicting, we introduce $c$, the probability that a pair of goals is incompatible:

\begin{itemize}
	\item $Pr(c=1 | g^1_p, g^2_p)=1$: Goals $g^1_p$ and $g^2_p$ are in complete conflict; they are mutually exclusive.
	\item $Pr(c=1 | g^1_p, g^2_p)=0$: Goals $g^1_p$ and $g^2_p$ are completely compatible and aligned.
\end{itemize}

For readability, we use $c(g^1_p, g^2_p)$ as shorthand for $Pr(c=1 | g^1_p, g^2_p)$.Since $c$ represents the probability of conflict between goals, it is a real number bounded in the range $[0,1]$. By construction, $c(g^x_p,g^x_p)=0$ and $c(g^x_p,g^0_p)=0$.

\textbf{3.3 \quad Goal Groups.} We define a goal group as a subset of agents that holds the same goal:

\begin{equation}
\mathcal{A}_g = \{a \in \mathcal{A} | h(a, g, j) \}
\end{equation}

By construction, $\mathcal{A} = \bigcup_g \mathcal{A}_g$.

\textbf{3.4 \quad Quantifying Misalignment.} In a single problem area, we can quantify misalignment as the probability that two randomly selected agents will hold incompatible goals:

\begin{equation}
\begin{aligned}
    Pr(1 | \mathcal{A}, p) := & \,\, Pr(a_1, a_2 \text{ selected randomly from } \mathcal{A}, \\ 
    & a_1 \neq a_2) \cdot c(g^1_p, g^2_p)
\end{aligned}
\end{equation}

The multiplication of the sampling and conflict components calculates the probability that any two randomly selected agents' goals will be in conflict. By doing this for all possible pairs of agents, we obtain an overall measure of misalignment in the population for a given problem area. The equation assumes the following constraints:
\begin{enumerate}
    \item Each agent holds no more than one goal per problem area.
    \item A lack of a goal ($g_0$) is not in conflict with any explicit goal.
    \item All agents are equally likely to be selected.
\end{enumerate}
We discuss the practical limitations of these constraints and how to remove them in Appendix A.
Note that the conflict function $c(\cdot,\cdot)$ allows for varying degrees of conflict between goals, rather than just binary conflict/no-conflict situations. Finally, the equation provides a single real scalar in the range $[0,1]$ representing the overall misalignment in the population for a specific problem area. The benefit of this approach is that it affords a simple representation capturing the real-world phenomena of ``strange bedfellows,'' where a pair of agents may be highly aligned in one specific area despite being highly misaligned in others.

\textbf{3.5 \quad Mutually Exclusive Goals.} To simplify our analysis, we can consider scenarios with mutually exclusive goals, meaning that each goal in $p$ completely conflicts with each other goal in $p$. This constraint allows for a more straightforward quantification of misalignment as follows:

\begin{equation} 
Pr(1|\mathcal{A},p) = \frac{\sum_{g \in \mathcal{G}}\sum_{g' \in \mathcal{G}, g' < g} 2|\mathcal{A}_g||\mathcal{A}_{g'}|}{|\mathcal{A}|(|\mathcal{A}|-1)} 
\end{equation}

Where $|\mathcal{A}_g|$ is the number of agents holding goal $g$ and $|\mathcal{A}|$ is the total number of agents. The alignment probability is then simply $Pr(0|\mathcal{A},p) = 1 - Pr(1 | \mathcal{A},p)$.

This equation allows us to derive a parametric quantity for misalignment of uniformly distributed agents in the range $[0,\frac{|\hat{\mathcal{G}}|-1}{|\hat{\mathcal{G}}|})$. See \S{5.1}, \S{5.2}, and \S{5.4} for experimental confirmation of this bound, and Appendices B.3 and B.4 for mathematical proof and interpretation.

\begin{algorithm}[tb]
\caption{\small Initialize World}
\begin{algorithmic}[1]
\REQUIRE $m = |\mathcal{P}|$, $n = |\mathcal{A}|$, $K = [|\mathcal{G}_1|,\ldots,|\mathcal{G}_m|]$
\REQUIRE Randomize, Range, Preset
\STATE world.$\mathcal{P} \leftarrow [p_1,\ldots,p_m]$
\FOR{$j = 1$ to $m$}
    \STATE world.$p_j$.$\mathcal{G} \leftarrow [g_1,\ldots,g_{K[j]}]$
    \FOR{$k = 1$ to $K[j]$}
        \FOR {$l = k+1$ to $K[j]$}
            \IF{Randomize.conflict = TRUE}
                \STATE world.$p_j$.c($g_k$,$g_l$) $\leftarrow$ \\ random(Range.conflict.min, .max)
            \ELSE
                \STATE world.$p_j$.c($g_k$,$g_l$) $\gets$ Preset.conflict[j][k][l]
            \ENDIF
        \ENDFOR
    \ENDFOR
\ENDFOR
\STATE world $\leftarrow$ ADD-AGENTS(world, Randomize, Range, Preset, $m$, $n$)
\RETURN world
\end{algorithmic}
\end{algorithm}

\textbf{3.6 \quad Overall Misalignment Across Problem Areas.} To find the misalignment of $\mathcal{A}$ across all problem areas, we take the arithmetic mean of $Pr(1|\mathcal{A}, p)$ across all $p$'s for that $\mathcal{A}$. Some problem areas are more important to us than others, and this importance changes depending on the agent. To aggregate misalignment accordingly, let each agent have a weight parameter $w$ for each problem area, such that 0 is the minimum amount an agent can care about a problem area, and 1 is the maximum amount. Then, multiply each problem area's misalignment by the geometric mean of the sampled agents' weights for that $p$. This gives us the following equation:

\begin{equation}
\begin{aligned}
    Pr(1 | \mathcal{A}, \mathcal{P}) := \\
     \frac{1}{|\mathcal{P}|} \sum_{p \in \mathcal{P}} & \Big( Pr(a_1, a_2 \text{ selected randomly from } \mathcal{A} \\
    & a_1 \neq a_2) \cdot c(g^1_p, g^2_p) \cdot \sqrt{w^1_p \cdot w^2_p} \Big)
\end{aligned}
\end{equation}

\begin{algorithm}[tb]
\caption{\small Create and add agents to world}
\begin{algorithmic}[1]
\REQUIRE world, Randomize, Range, Preset, $m$, $n$
\FOR{$i = 1$ to $n$}
    \STATE $a_i \leftarrow \{\}$
    \FOR{$j = 1$ to $m$}
        \IF{Randomize.goals = TRUE}
            \STATE $a_i.g^i_j \leftarrow$ random(world.$p_j$.$\mathcal{G}$)
        \ELSE
            \STATE $a_i.g^i_j \leftarrow$ Preset.goals[i][j]
        \ENDIF
        \IF{Randomize.weights = TRUE}
            \STATE $a_i.w^i_j \leftarrow$ random(Range.weights.min, .max)
        \ELSE
            \STATE $a_i.w^i_j \leftarrow$ Preset.weights[i][j]
        \ENDIF
    \ENDFOR
    \STATE world.$\mathcal{A} \leftarrow$ world.$\mathcal{A} \cup \{a_i\}$
\ENDFOR
\RETURN world
\end{algorithmic}
\end{algorithm}

\textbf{3.7 \quad Interpreting Misalignment Scores.} The misalignment score provides insights into the degree of goal conflict within a set of agents:

\begin{itemize}
    \item A score of 0 indicates perfect alignment (all agents share identical or fully compatible goals).
    \item A score approaching $\frac{|\hat{\mathcal{G}}|-1}{|\hat{\mathcal{G}}|}$ indicates high misalignment (goals are evenly distributed and in conflict).
    \item Intermediate scores suggest varying degrees of misalignment, which can be further analyzed by examining specific goal distributions and conflict levels.
\end{itemize}

This model provides a flexible framework for analyzing misalignment in complex scenarios involving multiple agents, problem areas, and goals. It can be applied to a wide range of situations, from social dynamics to AI alignment challenges.

\section{Simulation \& Experimental Setup}
To gain a deeper understanding of how different variables in our model influence misalignment, especially in larger-scale settings, we conducted a series of simulations. These empirical experiments complement our theoretical model and inform our later case studies.

We created abstract "worlds" with various configurations of problem areas, goals, agents, conflict scores, and importance weights. This approach allowed us to systematically explore the impact of changing one or more variables on overall misalignment scores.

Our simulation framework, outlined in Algorithms 1 and 2, initializes a world with specified parameters and populates it with agents. Each problem area has one or more goals (including a null goal with zero weight), and agents hold one goal per problem area. We treat an agent with zero weight in a problem area as equivalent to holding no goal in that area.

We conducted the following experiments, each focusing on different aspects of our model:

\begin{enumerate}
    \item \textbf{Varying Problem Areas}: We randomly assigned non-zero goals to agents, plotting results for different numbers of problem areas, each with three goals. All agents held their goals with maximum weight, and all goal pairs had maximum conflict values (Figure 1).
    
    \item \textbf{Varying Goals}: Similar to the first experiment, but we kept the number of problem areas constant and varied the number of goals (Figure 2).
    
    \item \textbf{Weight Sensitivity}: We investigated the effect of changing weights for a single goal group. We simulated 1000 agents, varying the weight of Goal 1 in Problem Area 1 for agents holding that goal, while keeping all other goals at maximum weight. We plotted this for 1 to 4 problem areas with 2 or 4 goals each (Figure 3).
    
    \item \textbf{Goal Distribution}: We plotted 1000 agents distributed deterministically among 2 or more goals, varying the proportion of agents assigned to goal 1 and distributing the remaining agents evenly to the remaining goals. All conflicts and weights were kept constant at 1 (Figure 4).
    
    \item \textbf{Conflict Levels}: We varied the number of goals in each problem area and plotted the results using different numbers of problem areas and conflict levels, while randomly assigning each agent's weights for their goals (Figure 5).
\end{enumerate}

For overall population misalignment across multiple problem areas, we used the arithmetic mean of area-specific misalignment scores. We allowed agents' weights across problem areas to sum to more than 1, as normalizing weights led to unintuitive results where misalignment scores depended more on the number of problem areas than on the agents' relative prioritization of each area. In these experiments, we excluded null goals in order to make the effects more clear.

\section{Results}
We present key findings from the five experiments that explore different aspects of misalignment dynamics.

\paragraph{5.1 \quad Varying Problem Areas.} Figure 1 illustrates how misalignment evolves as we increase the number of agents, with varying numbers of problem areas. The results reveal a consistent pattern across different numbers of problem areas. Initially, there is high variance in misalignment due to the random distribution of goals among a small number of agents. However, as the population size increases, the misalignment converges to a stable value. Notably, this convergence point is independent of the number of problem areas, because the overall misalignment is calculated as the arithmetic mean across all areas and each problem area has the same population misalignment. This finding demonstrates that our model captures the intuitive notion that larger populations tend to stabilize in their overall misalignment, even when they may have different early dynamics. It suggests that in complex multi-agent systems, the aggregate level of misalignment becomes more predictable as the number of agents grows and their goal distribution settles.

\paragraph{5.2 \quad Varying Goals.} Figure 2 shows that as the number of conflicting goals per problem area increases, so does the overall misalignment. For a problem area with with $\hat{\mathcal{G}}$ non-zero goals, the misalignment approaches $\frac{\hat{\mathcal{G}}-1}{\hat{\mathcal{G}}}$ as the number of agents grows. This pattern aligns with the intuitive understanding that a greater number of mutually exclusive goals leads to higher potential for misalignment within a population. The result highlights our model's ability to capture the complexity of multi-goal scenarios, reflecting the increased potential for conflict as the number of distinct goals grows. It suggests that in real-world situations, systems with a higher number of competing objectives are likely to exhibit greater levels of misalignment, showing the importance of goal prioritization and conflict resolution in multi-agent environments.

\begin{figure}[t]
\centering
\includegraphics[width=0.85\columnwidth]{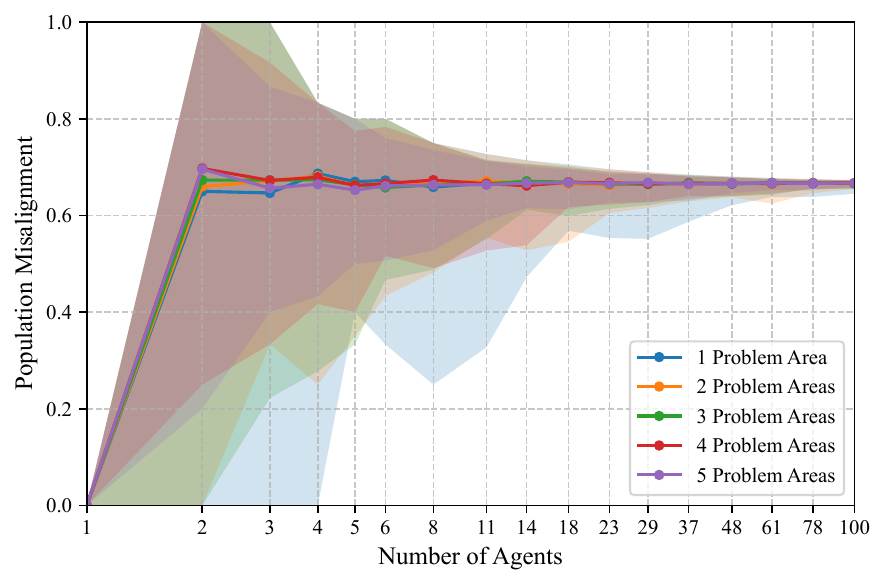}
\caption{\small \textbf{Varying Problem Areas.} Random goal assignment, 3 goals per problem area, 100 runs per data point, max weights mutually exclusive and conflict values. Null goals disallowed.}
\end{figure}

\begin{figure}[t]
\centering
\includegraphics[width=0.85\columnwidth]{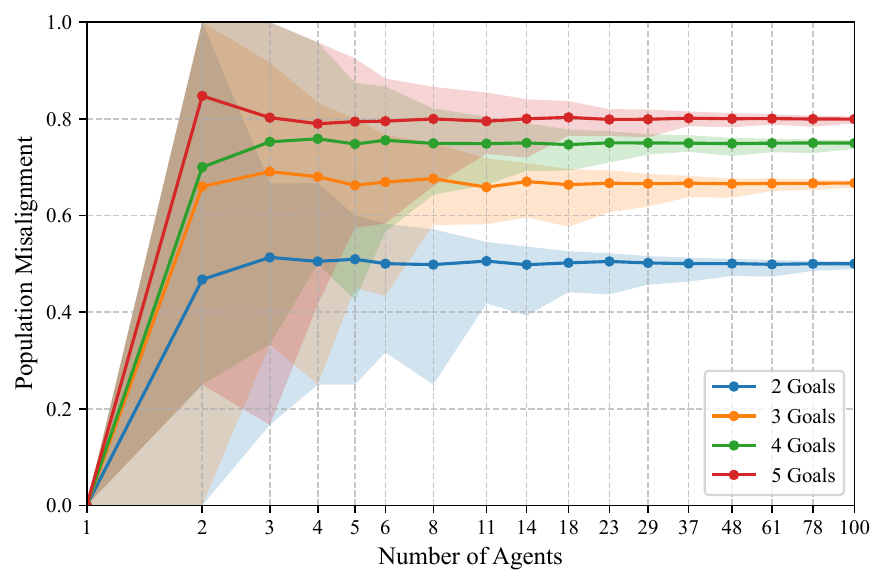}
\caption{\small \textbf{Varying Goals.} Random goal assignment, 4 problem areas, 100 runs per data point, max weight and conflict values. Null goals disallowed.}
\end{figure}

\paragraph{5.3 \quad Weight Sensitivity.} Figure 3 examines the effect of changing weights for a single goal group in one problem area, while keeping other weights at maximum. When the weight of one goal group is reduced to zero in a single problem area, it effectively eliminates misalignment for that specific area. This aspect of the model captures scenarios where some objectives are irrelevant or unimportant to a subset of agents. As more problem areas are added to the simulation, the impact of weight changes in a single area diminishes, reflecting the model's capability to balance multiple concerns. Interestingly, when the number of goals per problem area is increased to four, the misalignment starts at a higher level and approaches 0.75, consistent with the $\frac{\hat{\mathcal{G}}-1}{\hat{\mathcal{G}}}$ pattern observed in the previous experiment. These findings show that our model is sensitive to goal weights and can capture nuanced interactions between problem areas. They suggest that in complex systems, the relative importance assigned to different goals can significantly influence overall alignment, and that this effect is modulated by the number of problem areas considered.

\begin{figure}[t]
\centering
\includegraphics[width=0.85\columnwidth]{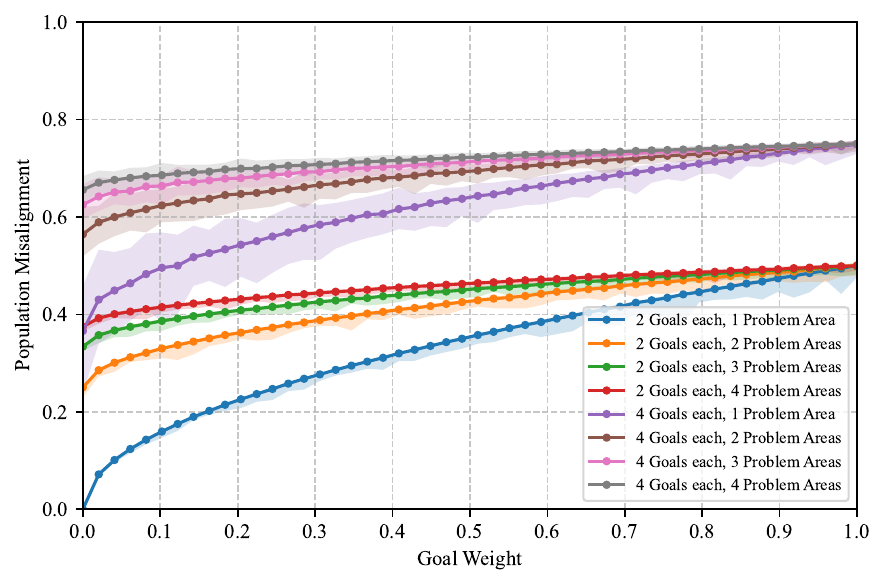}
\caption{\small \textbf{Weight Sensitivity.} 100 agents, 100 runs per data point, maximum conflict, random goal assignment. Null goals disallowed.}
\end{figure}

\begin{figure}[t]
\centering
\includegraphics[width=0.85\columnwidth]{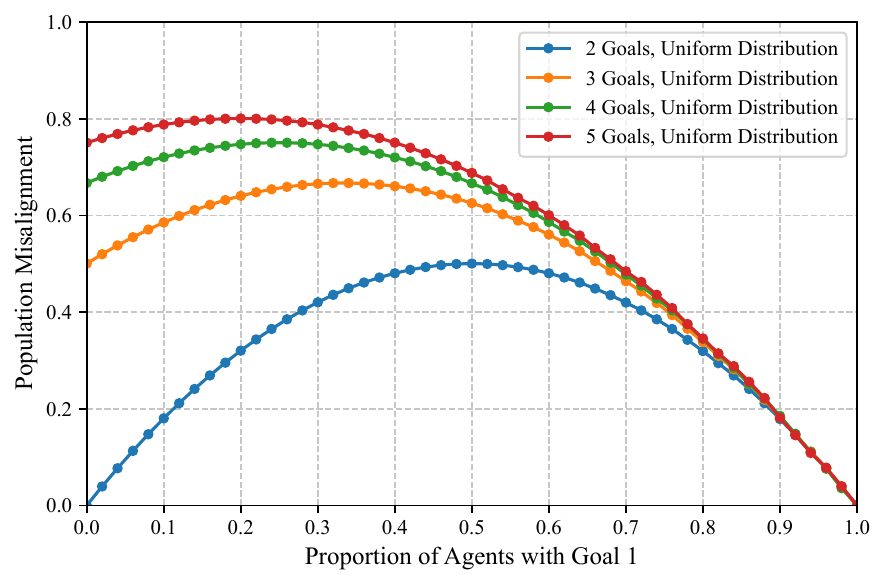}
\caption{\small \textbf{Goal Distribution.} 1000 agents, 1 problem area, max conflict and weight. Null goals disallowed.}
\end{figure}

\begin{figure}[t]
\centering
\includegraphics[width=0.85\columnwidth]{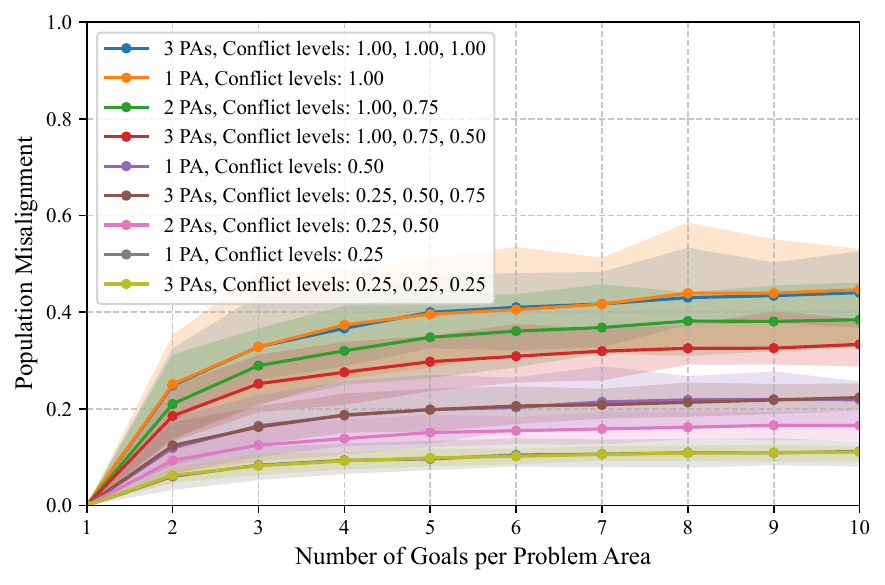}
\caption{\small \textbf{Conflict Levels.} 120 agents, 100 runs, random weight range (0.25, 0.75). Null goals disallowed.}
\end{figure}

\paragraph{5.4 \quad Goal Distribution.} Figure 4 shows how the dynamics of misalignment change as we vary the proportion of agents assigned to different goals. Notably, misalignment reaches its peak when goals are evenly distributed among agents. Specifically, for a problem area with $\hat{\mathcal{G}}$ non-zero goals, the maximum misalignment occurs when the proportion of agents assigned to any single goal is $\frac{1}{\hat{\mathcal{G}}}$. As the population becomes more homogeneous in its goals - that is, as a larger proportion of agents adopt the same goal - the overall misalignment decreases. As the Goal 1 group shrinks, agents are redistributed, so the leftmost value of each curve is the peak value of the curve with one less goal. This pattern holds regardless of the total number of goals, though the peak misalignment value increases with the number of available goals. The results suggest (intuitively) that in real-world scenarios, divided populations with few common objectives are more misaligned than to those with a more dominant consensus. %

\paragraph{5.5 \quad Conflict Levels.} Figure 5 explores the effects of varying conflict levels across different numbers of problem areas and goals. Each problem area is assigned a single conflict value for all of its goal pairs.

As expected, misalignment increases with both the number of goals and the level of conflict between goals. However, the effect of multiple problem areas is more nuanced, depending on the average conflict across those areas rather than their individual values. Notably, scenarios with the same average conflict level converge to similar misalignment values, regardless of how that average is achieved across different problem areas. This finding suggests that our model effectively captures the aggregate effect of conflicts across multiple domains, providing a holistic view of misalignment. More broadly, this demonstrates our model's ability to handle intricate scenarios with varying conflict levels across multiple problem areas, offering a sophisticated tool for analyzing misalignment in multi-faceted, real-world situations where conflicts may be unevenly distributed across different domains of interaction.

Together, our experiments validate key properties of our misalignment model, demonstrating its ability to capture intuitive aspects of multi-agent alignment while revealing nontrivial dynamics in complex scenarios. With this solid foundation, we now apply our model to realistic case studies. 

\section{Case Studies}
To demonstrate the robustness and applicability of our misalignment model, we present two distinct case studies: a shopping recommender system and an autonomous vehicle pre-collision scenario.

\textbf{6.1 \quad Shopping Recommender System.}
Consider an AI-based recommender system employed by a hybrid "big-box" retailer with both online and brick-and-mortar presence. This system mediates between the retailer and its customers, potentially leading to varying degrees of alignment or misalignment depending on the specific shopping context.

Figure 6 illustrates two scenarios:

\begin{enumerate}
    \item \textbf{Grocery Shopping:} A customer uses the retailer's mobile app to quickly order groceries for curbside pickup. The recommender system suggests familiar items and meal combinations, saving time and providing convenience. This scenario demonstrates alignment between customer, retailer, and recommender system goals.
    
    \item \textbf{Impulse Purchasing:} Late at night, the same customer is led by the recommender system to purchase unnecessary clearance holiday items. This results in guilt, wasted money and time, and refunded items, demonstrating misalignment between all parties' interests.
\end{enumerate}

\begin{figure}[t] 
    \centering
    \includegraphics[width=0.9\columnwidth]{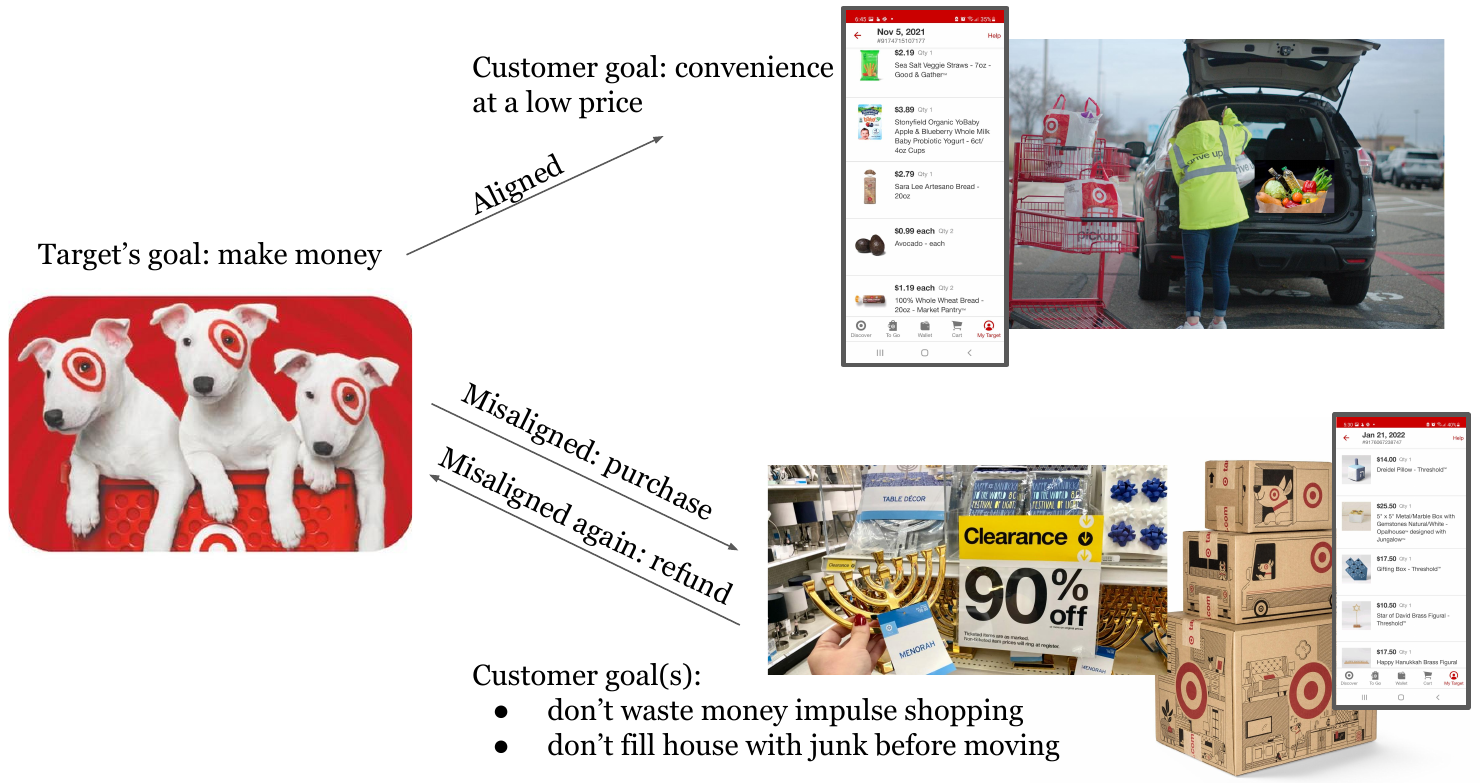}
    \caption{\small Shopping Recommender System scenarios. Top: Aligned grocery shopping. Bottom: Misaligned impulse purchasing.}
\end{figure}

Let's map this scenario to our model:

\begin{itemize}
    \item Agents: $\mathcal{A} = \{c, R, R^{S}\}$, where $c$ is the customer, $R$ is the retailer, and $R^{S}$ is the recommender system.\footnote{Though we only use one instance of each agent here, it is easy to add agents to any category and recalculate accordingly.}
    \item Problem Areas: $\mathcal{P} = \{p_f, p_h\}$, where $p_f$ is food and grocery shopping and $p_h$ is household item shopping.
\end{itemize}

\begin{table}[h]
\centering
\setlength{\tabcolsep}{3pt}
\resizebox{0.75\columnwidth}{!}{
\begin{tabular}{|l|l|l|}
\hline
\rule{0pt}{2.2ex}Goal & Description & Weight ($w$) \\[0.5ex]
\hline \hline
\rule{0pt}{2.2ex}$g^{c}_{p_f}$ & Convenience at low price & $w^{c}_{p_f} = 0.8$ \\[0.5ex]
\hline
\rule{0pt}{2.2ex}$g^{c}_{p_h}$ & Avoid impulse buying & $w^{c}_{p_h} = 0.6$ \\[0.5ex]
\hline
\rule{0pt}{2.5ex}$g^{R}_{p_f}$ & Increase net profits & $w^{R}_{p_f} = 0.9$ \\[0.5ex]
\hline
\rule{0pt}{2.5ex}$g^{R}_{p_h}$ & Move inventory & $w^{R}_{p_h} = 0.7$ \\[0.5ex]
\hline
\rule{0pt}{2.5ex}$g^{R^{S}}_{p_f}$ & Maximize checkout value & $w^{R^{S}}_{p_f} = 1.0$ \\[0.5ex]
\hline
\rule{0pt}{2.5ex}$g^{R^{S}}_{p_h}$ & Maximize checkout value & $w^{R^{S}}_{p_h} = 1.0$ \\[0.5ex]
\hline
\end{tabular}
}
\caption{\small Shopping scenario goals and weights}
\end{table}

\begin{table}[ht]
\centering
\resizebox{0.9\columnwidth}{!}{ %
\begin{tabular}{cc}
\multicolumn{2}{c}{\textbf{}} \\
\textbf{Food:} & \textbf{Household:} \\[0.5em]
$\begin{array}{c|ccc}
 & g^c & g^R & g^{R^S} \\
\hline
g^c & 0 & 0.1 & 0.1 \\
g^R &    & 0 & 0.1 \\
g^{R^S} &    &   & 0 \\
\end{array}$
&
$\begin{array}{c|ccc}
 & g^c & g^R & g^{R^S} \\
\hline
g^c & 0 & 0.5 & 0.9 \\
g^R &   & 0 & 0.3 \\
g^{R^S} &   &   & 0 \\
\end{array}$
\end{tabular}
}
\caption{\small Lower triangular matrices representing goal conflicts in two problem areas. \textbf{Left:} Conflict matrix for problem area $p_f$ (Food). \textbf{Right:} Conflict matrix for problem area $p_h$ (Household). The entries represent the conflict values between goals $g^c$, $g^R$, and $g^{R^S}$, with diagonal entries set to 0, indicating no conflict with themselves.}
\end{table}

We can now calculate misalignment in each problem area:

\begin{equation}
Pr(1 | \mathcal{A}, p) = \frac{\sum_{a_1, a_2 \in \mathcal{A}} c(g_{a_1}, g_{a_2}) \cdot \sqrt{w_{a_1} \cdot w_{a_2}}}{|\mathcal{A}|(|\mathcal{A}|-1)} 
\end{equation}

And the overall misalignment:

\begin{equation}
Pr(1 | \mathcal{A}, \mathcal{P}) = \frac{1}{|\mathcal{P}|} \sum_{p \in \mathcal{P}} Pr(1 | \mathcal{A}, p)
\end{equation}

This results in the following scores: $Pr(1 | \mathcal{A}, p_f) = 0.09$, $Pr(1 | \mathcal{A}, p_h) = 0.42$, $Pr(1 | \mathcal{A}, \mathcal{P}) = 0.26$. This lines up with our intuition that all three agents are much more aligned w/r/t the grocery area than the household area. Overall, this case study demonstrates how our model can capture the nuanced misalignment in e-commerce scenarios, where goals may align in one context (grocery shopping) but conflict in another (impulse shopping), and where recommender system incentives may thus lead to subtle reward hacking.

\textbf{6.2 \quad Autonomous Vehicle Pre-collision Scenario.}
We analyze scenario 17 from the CARLA (Car Learning to Act) challenge: "Obstacle avoidance without prior action."~\cite{dosovitskiy2017carla} This scenario involves an autonomous vehicle encountering an unexpected pedestrian and needing to perform an emergency maneuver.

\begin{figure}[ht] 
    \centering
    \includegraphics[width=0.7\columnwidth]{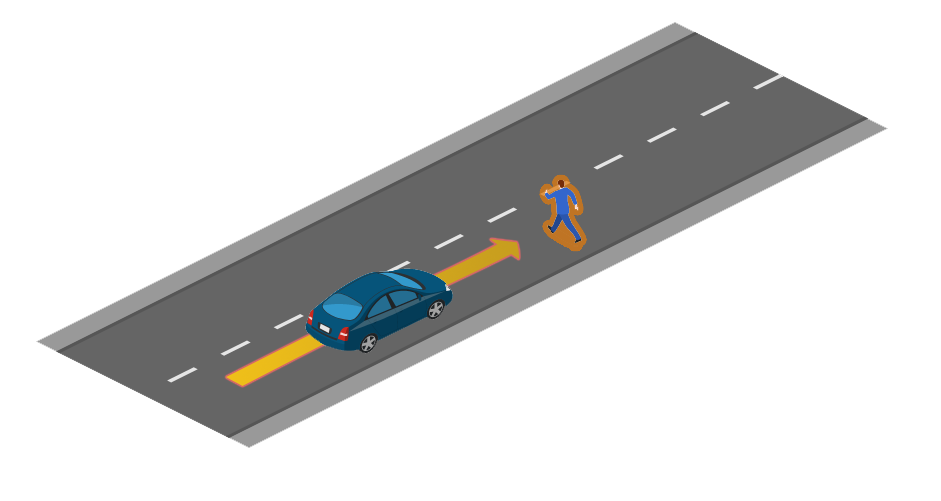}
    \caption{\small CARLA scenario 17 - Autonomous vehicle encountering a pedestrian~\cite{dosovitskiy2017carla}.}
\end{figure}

Figure 7 illustrates the scenario where the autonomous vehicle must make rapid decisions to avoid a collision while considering various factors such as passenger safety, pedestrian safety, and traffic rules. In this case:
\begin{itemize}
    \item Agents: $\mathcal{A} = \{a_{v}, a_{h}\}$, where $a_{v}$ is the autonomous vehicle and $a_{h}$ is the human pedestrian.
    \item Problem Areas: $\mathcal{P} = \{p_1, ..., p_8\}$, corresponding to penalized categories of traffic events in the CARLA challenge.
\end{itemize}

In Table 3, goals for both agents in all problem areas are to avoid the described event (e.g., collision, traffic violation). Weights for $a_{v}$ are derived from CARLA penalty coefficients for each event, and estimated for $a_{h}$. While both agents prefer to avoid each event, we assume that pedestrian and AV goals may conflict in \textit{how} they wish to avoid incident, and simulate this as some level of conflict between $g^{v}_{p}$ and $g^{h}_{p}$ for each $p$.

\begin{table}[ht]
\centering
\setlength{\tabcolsep}{3pt}
\resizebox{1\columnwidth}{!}{
\begin{tabular}{|l|l|l|l|}
\hline
\rule{0pt}{2.2ex}Goal & Description & \multicolumn{2}{c|}{Weights} \\[0.2ex]
\hline
\hline
\rule{0pt}{2.5ex}$g^{v|h}_{p1}$ & No pedestrian collision & $w^{v}_{p1} = 0.50$ & $w^{h}_{p1} = 0.99$ \\[0.5ex]
\hline
\rule{0pt}{2.5ex}$g^{v|h}_{p2}$ & No vehicle collision & $w^{v}_{p2} = 0.40$ & $w^{h}_{p2} = 0.15$ \\[0.5ex]
\hline
\rule{0pt}{2.5ex}$g^{v|h}_{p3}$ & No static object collision & $w^{v}_{p3} = 0.35$ & $w^{h}_{p3} = 0.15$ \\[0.5ex]
\hline
\rule{0pt}{2.5ex}$g^{v|h}_{p4}$ & No red light violation & $w^{v}_{p4} = 0.30$ & $w^{h}_{p4} = 0.05$ \\[0.5ex]
\hline
\rule{0pt}{2.5ex}$g^{v|h}_{p5}$ & No stop sign violation & $w^{v}_{p5} = 0.20$ & $w^{h}_{p5} = 0.05$ \\[0.5ex]
\hline
\rule{0pt}{2.5ex}$g^{v|h}_{p6}$ & No route blockage & $w^{v}_{p6} = 0.30$ & $w^{h}_{p6} = 0.05$ \\[0.5ex]
\hline
\rule{0pt}{2.5ex}$g^{v|h}_{p7}$ & Keep appropriate speed & $w^{v}_{p7} = 0.30$ & $w^{h}_{p7} = 0.01$ \\[0.5ex]
\hline
\rule{0pt}{2.5ex}$g^{v|h}_{p8}$ & No yield violation & $w^{v}_{p8} = 0.30$ & $w^{h}_{p8} = 0.05$ \\[0.5ex]
\hline
\end{tabular}
}
\caption{\small Autonomous vehicle case. "$v|h$" means each goal is held by both $v$ehicle and $h$uman (pedestrian), though their weights vary.}
\end{table}

\begin{figure}[ht]
\centering
\includegraphics[width=0.85\columnwidth]{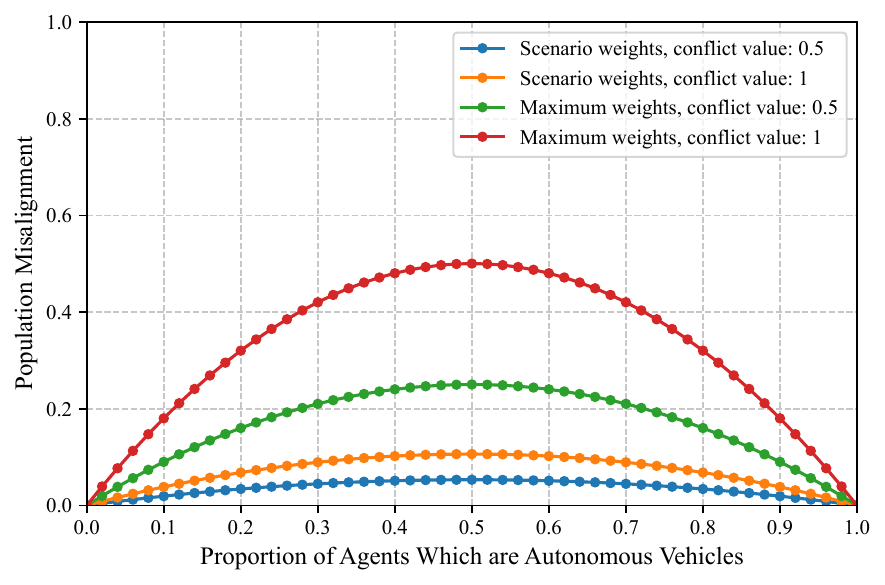} %
\caption{\small CARLA scenario, 1000 agents, assuming goals conflict.} %
\end{figure}

To obtain population misalignment scores, we simulate a world with 1000 agents that are pedestrians or autonomous vehicles (see Figure 8). As in the goal distribution experiment (\S{5.4}), a population with opposing goals has the highest misalignment when those goals are most evenly distributed. We note that even with $c(g^v_p, g^h_p)=1$, the high number of problem areas with low weights results in low misalignment at almost all times. We add curves with maximum weights to confirm that this is indeed an effect of the low weights.

\section{Discussion}
Our approach enables modeling misalignment in diverse contexts, from international relations to family disagreements. It captures a variety of real-world phenomena, such as the ``strange bedfellows'' effect (see \S{3.4}) and reward hacking (see \S{6.1}), and allows for misalignment that varies across different populations and problem areas, providing a comprehensive framework for understanding alignment dynamics in diverse contexts. By addressing the fundamental challenge of defining alignment in a world where humans---and not just AI agents---are frequently misaligned, our model affords the analysis of complex scenarios ranging from localized disputes to global conflicts..

\textbf{7.1 \quad Implications for AI Risk.} In the context of AI risk analysis, our work contributes to the understanding of existential risks posed by misaligned AI \cite{Avin18Classifying,LongTermTraj,Precipice,bucknall2022current}. By reframing misalignment as primarily a human-centered problem, we align with recent research on AI's impact on human decision-making and societal structures \cite{russell2019human,Colemane2025764118,SegerEpistemicSecurity}. Our model provides a tool for analyzing potential AI-powered conflicts and their broader implications \cite{SIPRI19Vol1,JohnsonAI,HerbertLin,Maas2022MilitaryAI}, highlighting the complexity of aligning AI with human values.

Furthermore, as a framework to measure socio-technical misalignment, our approach could lead to better training procedures that account for value pluralism and more effective regulatory frameworks for ensuring AI systems achieve appropriate levels of alignment across different contexts and populations. See Appendix C for more detail.

\textbf{7.2 \quad Limitations.} Our model does not address whether an agent's actions have positive or negative outcomes for the agent itself. The question of how an agent's goals could be learned remains open, although progress has been made by others in this space \cite{brownvaluealignmentverification}. We discuss other assumptions and restrictions in Appendix A.

\section{Conclusion}

We introduce a novel misalignment model which advances the understanding of AI alignment by focusing on multi-agent settings and diverse problem areas. Building on a computational social science model of contention in human populations \cite{jang2017modeling}, we introduce a framework that captures the nuanced, multi-faceted nature of alignment in complex sociotechnical systems. Our simulations demonstrate that our model's misalignment scores accurately reflect changes in key variables, providing a robust tool for analyzing multi-agent alignment scenarios. This approach offers a more nuanced and realistic understanding of misalignment than simple binary classifications or global numeric values, better reflecting the complexities of 
real-world interactions between humans and AI systems.

\textbf{8.1 \quad Broader Impacts.} Our approach encourages AI safety and alignment researchers to avoid reductionist traps, such as attempting to align AI narrowly with individuals or humanity as a whole, or adopting an overly techno-solutionist mindset. We hope to spark conversation about the sociotechnical aspects of the alignment problem and the need for interdisciplinary collaboration in addressing these challenges. We discuss further impacts in Appendix C.

\textbf{8.2 \quad Future Work.} Future research could extend our model to entities beyond humans and AI, including social constructs like nation-states and corporations (alluded to in \S{6.1}), as well as biological entities from cellular interactions to ecosystems. This expansion could provide insights into alignment dynamics in diverse complex systems.

Challenges remain in precisely defining problem areas, particularly when they involve combinations or gestalts of sub-areas, have mutual dependencies, or when selecting relevant states from potentially infinite possibilities. Addressing these challenges to produce more rigorous definitions of problem areas is an important direction for future work.

Finally, applying our model to existing multi-agent simulations, such as those in MeltingPot~\cite{leibo2021meltingpot}, DeepMind's Concordia~\cite{vezhnevets2023generative}, and other population simulations~\cite{piatti2024cooperatecollapseemergencesustainable,foxabbott-defining-2023} could yield valuable insights into the dynamics of misalignment in complex multi-agent systems.

\section*{Acknowledgements} 
This material is based upon work supported in part by the NSF Program on Fairness in AI in Collaboration with Amazon under Award IIS-2147305. Any opinion, findings, and conclusions or recommendations expressed in this material are those of the author(s) and do not necessarily reflect the views of the National Science Foundation or Amazon.

\bibliography{aaai25}

\begin{thebibliography}{30}
\providecommand{\natexlab}[1]{#1}

\bibitem[{Avin et~al.(2018)Avin, Wintle, Weitzdörfer, {Ó hÉigeartaigh}, Sutherland, and Rees}]{Avin18Classifying}
Avin, S.; Wintle, B.~C.; Weitzdörfer, J.; {Ó hÉigeartaigh}, S.~S.; Sutherland, W.~J.; and Rees, M.~J. 2018.
\newblock Classifying global catastrophic risks.
\newblock \emph{Futures}, 102: 20--26.

\bibitem[{Bak-Coleman et~al.(2021)Bak-Coleman, Alfano, Barfuss, Bergstrom, Centeno, Couzin, Donges, Galesic, Gersick, Jacquet, Kao, Moran, Romanczuk, Rubenstein, Tombak, Van~Bavel, and Weber}]{Colemane2025764118}
Bak-Coleman, J.~B.; Alfano, M.; Barfuss, W.; Bergstrom, C.~T.; Centeno, M.~A.; Couzin, I.~D.; Donges, J.~F.; Galesic, M.; Gersick, A.~S.; Jacquet, J.; Kao, A.~B.; Moran, R.~E.; Romanczuk, P.; Rubenstein, D.~I.; Tombak, K.~J.; Van~Bavel, J.~J.; and Weber, E.~U. 2021.
\newblock Stewardship of global collective behavior.
\newblock \emph{Proceedings of the National Academy of Sciences}, 118(27).

\bibitem[{Baum et~al.(2019)Baum, Armstrong, Ekenstedt, H\"aggstr\"om, Hanson, Kuhlemann, Maas, Miller, Salmela, Sandberg, Sotala, Torres, Turchin, and Yampolskiy}]{LongTermTraj}
Baum, S.~D.; Armstrong, S.; Ekenstedt, T.; H\"aggstr\"om, O.; Hanson, R.; Kuhlemann, K.; Maas, M.~M.; Miller, J.~D.; Salmela, M.; Sandberg, A.; Sotala, K.; Torres, P.; Turchin, A.; and Yampolskiy, R.~V. 2019.
\newblock Long-term trajectories of human civilization.
\newblock \emph{Foresight}, 21(1): 53--83.

\bibitem[{Boulanin et~al.(2019)Boulanin, Avin, Sauer, Borrie, Scheftelowitsch, Bronk, Stoutland, Hagstr\"om, Topychkanov, Horowitz, Kaspersen, King, Amadae, and Rickli}]{SIPRI19Vol1}
Boulanin, V.; Avin, S.; Sauer, F.; Borrie, J.; Scheftelowitsch, D.; Bronk, J.; Stoutland, P.~O.; Hagstr\"om, M.; Topychkanov, P.; Horowitz, M.~C.; Kaspersen, A.; King, C.; Amadae, S.; and Rickli, J.-M. 2019.
\newblock The Impact of Artificial Intelligence on Strategic Stability and Nuclear Risk, Volume I, Euro-Atlantic Perspectives.
\newblock Technical report, SIPRI.

\bibitem[{Brown et~al.(2021)Brown, Schneider, Dragan, and Niekum}]{brownvaluealignmentverification}
Brown, D.~S.; Schneider, J.; Dragan, A.; and Niekum, S. 2021.
\newblock Value Alignment Verification.
\newblock In Meila, M.; and Zhang, T., eds., \emph{Proceedings of the 38th International Conference on Machine Learning}, volume 139 of \emph{Proceedings of Machine Learning Research}, 1105--1115. PMLR.

\bibitem[{Bucknall and Dori-Hacohen(2022)}]{bucknall2022current}
Bucknall, B.~S.; and Dori-Hacohen, S. 2022.
\newblock Current and Near-Term AI as a Potential Existential Risk Factor.
\newblock In \emph{Proceedings of the 2022 AAAI/ACM Conference on AI, Ethics, and Society}, AIES '22, 119–129. New York, NY, USA: Association for Computing Machinery.
\newblock ISBN 9781450392471.

\bibitem[{Chan et~al.(2023)Chan, Salganik, Markelius, Pang, Rajkumar, Krasheninnikov, Langosco, He, Duan, Carroll et~al.}]{chan2023harms}
Chan, A.; Salganik, R.; Markelius, A.; Pang, C.; Rajkumar, N.; Krasheninnikov, D.; Langosco, L.; He, Z.; Duan, Y.; Carroll, M.; et~al. 2023.
\newblock Harms from increasingly agentic algorithmic systems.
\newblock In \emph{Proceedings of the 2023 ACM Conference on Fairness, Accountability, and Transparency}, 651--666.

\bibitem[{Christian(2020)}]{christian2020alignment}
Christian, B. 2020.
\newblock \emph{The Alignment Problem: Machine Learning and Human Values}.
\newblock WW Norton \& Company.

\bibitem[{Critch and Krueger(2020)}]{critch2020ai}
Critch, A.; and Krueger, D. 2020.
\newblock {AI} Research Considerations for Human Existential Safety {(ARCHES)}.
\newblock \emph{CoRR}, abs/2006.04948.

\bibitem[{Dori-Hacohen et~al.(2021)Dori-Hacohen, Sung, Chou, and Lustig-Gonzalez}]{dorihacohen2021restoring}
Dori-Hacohen, S.; Sung, K.; Chou, J.; and Lustig-Gonzalez, J. 2021.
\newblock \emph{Restoring Healthy Online Discourse by Detecting and Reducing Controversy, Misinformation, and Toxicity Online}, 2627--2628.
\newblock New York, NY, USA: Association for Computing Machinery.
\newblock ISBN 9781450380379.

\bibitem[{Dosovitskiy et~al.(2017)Dosovitskiy, Ros, Codevilla, Lopez, and Koltun}]{dosovitskiy2017carla}
Dosovitskiy, A.; Ros, G.; Codevilla, F.; Lopez, A.; and Koltun, V. 2017.
\newblock CARLA: An open urban driving simulator.
\newblock In \emph{Conference on robot learning}, 1--16. PMLR.

\bibitem[{Foxabbott et~al.(2023)Foxabbott, Deverett, Senft, Dower, and Hammond}]{foxabbott-defining-2023}
Foxabbott, J.; Deverett, S.; Senft, K.; Dower, S.; and Hammond, L. 2023.
\newblock Defining and {Mitigating} {Collusion} in {Multi}-{Agent} {Systems}.
\newblock In \emph{Multi-{Agent} {Security} {Workshop} @ {NeurIPS}'23}.

\bibitem[{Gabriel(2020)}]{gabrielartificial2020}
Gabriel, I. 2020.
\newblock Artificial {Intelligence}, {Values}, and {Alignment}.
\newblock \emph{Minds and Machines}, 30(3): 411--437.

\bibitem[{Gabriel et~al.(2024)Gabriel, Manzini, Keeling, Hendricks, Rieser, Iqbal, Tomašev, Ktena, Kenton, Rodriguez, El-Sayed, Brown, Akbulut, Trask, Hughes, Bergman, Shelby, Marchal, Griffin, Mateos-Garcia, Weidinger, Street, Lange, Ingerman, Lentz, Enger, Barakat, Krakovna, Siy, Kurth-Nelson, McCroskery, Bolina, Law, Shanahan, Alberts, Balle, de~Haas, Ibitoye, Dafoe, Goldberg, Krier, Reese, Witherspoon, Hawkins, Rauh, Wallace, Franklin, Goldstein, Lehman, Klenk, Vallor, Biles, Morris, King, y~Arcas, Isaac, and Manyika}]{gabriel2024ethicsadvancedaiassistants}
Gabriel, I.; Manzini, A.; Keeling, G.; Hendricks, L.~A.; Rieser, V.; Iqbal, H.; Tomašev, N.; Ktena, I.; Kenton, Z.; Rodriguez, M.; El-Sayed, S.; Brown, S.; Akbulut, C.; Trask, A.; Hughes, E.; Bergman, A.~S.; Shelby, R.; Marchal, N.; Griffin, C.; Mateos-Garcia, J.; Weidinger, L.; Street, W.; Lange, B.; Ingerman, A.; Lentz, A.; Enger, R.; Barakat, A.; Krakovna, V.; Siy, J.~O.; Kurth-Nelson, Z.; McCroskery, A.; Bolina, V.; Law, H.; Shanahan, M.; Alberts, L.; Balle, B.; de~Haas, S.; Ibitoye, Y.; Dafoe, A.; Goldberg, B.; Krier, S.; Reese, A.; Witherspoon, S.; Hawkins, W.; Rauh, M.; Wallace, D.; Franklin, M.; Goldstein, J.~A.; Lehman, J.; Klenk, M.; Vallor, S.; Biles, C.; Morris, M.~R.; King, H.; y~Arcas, B.~A.; Isaac, W.; and Manyika, J. 2024.
\newblock The Ethics of Advanced AI Assistants.
\newblock arXiv:2404.16244.

\bibitem[{Jang, Dori-Hacohen, and Allan(2017)}]{jang2017modeling}
Jang, M.; Dori-Hacohen, S.; and Allan, J. 2017.
\newblock Modeling controversy within populations.
\newblock In \emph{Proceedings of the ACM SIGIR International Conference on Theory of Information Retrieval}, 141--149.

\bibitem[{Ji et~al.(2024)Ji, Qiu, Chen, Zhang, Lou, Wang, Duan, He, Zhou, Zhang, Zeng, Ng, Dai, Pan, O'Gara, Lei, Xu, Tse, Fu, McAleer, Yang, Wang, Zhu, Guo, and Gao}]{ji2024ai}
Ji, J.; Qiu, T.; Chen, B.; Zhang, B.; Lou, H.; Wang, K.; Duan, Y.; He, Z.; Zhou, J.; Zhang, Z.; Zeng, F.; Ng, K.~Y.; Dai, J.; Pan, X.; O'Gara, A.; Lei, Y.; Xu, H.; Tse, B.; Fu, J.; McAleer, S.; Yang, Y.; Wang, Y.; Zhu, S.-C.; Guo, Y.; and Gao, W. 2024.
\newblock AI Alignment: A Comprehensive Survey.
\newblock arXiv:2310.19852.

\bibitem[{Johnson(2019)}]{JohnsonAI}
Johnson, J. 2019.
\newblock Artificial intelligence \& future warfare: implications for international security.
\newblock \emph{Defense \& Security Analysis}, 35(2): 147--169.

\bibitem[{Johnson et~al.(2022)Johnson, Pistilli, Men{\'e}dez-Gonz{\'a}lez, Duran, Panai, Kalpokiene, and Bertulfo}]{johnson2022ghost}
Johnson, R.~L.; Pistilli, G.; Men{\'e}dez-Gonz{\'a}lez, N.; Duran, L. D.~D.; Panai, E.; Kalpokiene, J.; and Bertulfo, D.~J. 2022.
\newblock The Ghost in the Machine has an American accent: value conflict in GPT-3.
\newblock \emph{arXiv preprint arXiv:2203.07785}.

\bibitem[{Kapoor et~al.(2024)Kapoor, Stroebl, Siegel, Nadgir, and Narayanan}]{kapoor2024aiagentsmatter}
Kapoor, S.; Stroebl, B.; Siegel, Z.~S.; Nadgir, N.; and Narayanan, A. 2024.
\newblock AI agents that matter.
\newblock \emph{arXiv preprint arXiv:2407.01502}.

\bibitem[{Lazar and Nelson(2023)}]{lazar2023ai}
Lazar, S.; and Nelson, A. 2023.
\newblock AI safety on whose terms?

\bibitem[{Leibo et~al.(2021)Leibo, Due{\~n}ez-Guzman, Vezhnevets, Agapiou, Sunehag, Koster, Matyas, Beattie, Mordatch, and Graepel}]{leibo2021meltingpot}
Leibo, J.~Z.; Due{\~n}ez-Guzman, E.~A.; Vezhnevets, A.; Agapiou, J.~P.; Sunehag, P.; Koster, R.; Matyas, J.; Beattie, C.; Mordatch, I.; and Graepel, T. 2021.
\newblock Scalable Evaluation of Multi-Agent Reinforcement Learning with Melting Pot.
\newblock In Meila, M.; and Zhang, T., eds., \emph{Proceedings of the 38th International Conference on Machine Learning}, volume 139 of \emph{Proceedings of Machine Learning Research}, 6187--6199. PMLR.

\bibitem[{Leike et~al.(2018)Leike, Krueger, Everitt, Martic, Maini, and Legg}]{leike2018scalable}
Leike, J.; Krueger, D.; Everitt, T.; Martic, M.; Maini, V.; and Legg, S. 2018.
\newblock Scalable agent alignment via reward modeling: a research direction.
\newblock ArXiv:1811.07871 [cs, stat].

\bibitem[{Lin(2019)}]{HerbertLin}
Lin, H. 2019.
\newblock The existential threat from cyber-enabled information warfare.
\newblock \emph{Bulletin of the Atomic Scientists}, 75(4): 187--196.

\bibitem[{Maas, Matteuci, and Cooke(2022)}]{Maas2022MilitaryAI}
Maas, M.~M.; Matteuci, K.; and Cooke, D. 2022.
\newblock Military Artificial Intelligence as Contributor to Global Catastrophic Risk.
\newblock In \emph{Cambridge Conference on Catastrophic Risks 2020}.
\newblock Draft available at \url{https://papers.ssrn.com/sol3/papers.cfm?abstract_id=4115010}.

\bibitem[{Ord(2020)}]{Precipice}
Ord, T. 2020.
\newblock \emph{{The Precipice: Existential Risk and the Future of Humanity}}.
\newblock Hachette Books.
\newblock ISBN 978-0316484916.

\bibitem[{Piatti et~al.(2024)Piatti, Jin, Kleiman-Weiner, Schölkopf, Sachan, and Mihalcea}]{piatti2024cooperatecollapseemergencesustainable}
Piatti, G.; Jin, Z.; Kleiman-Weiner, M.; Schölkopf, B.; Sachan, M.; and Mihalcea, R. 2024.
\newblock Cooperate or Collapse: Emergence of Sustainable Cooperation in a Society of LLM Agents.
\newblock arXiv:2404.16698.

\bibitem[{Russell(2019)}]{russell2019human}
Russell, S. 2019.
\newblock \emph{Human Compatible: {A}rtificial {I}ntelligence and the {P}roblem of {C}ontrol}.
\newblock Penguin.

\bibitem[{Seger et~al.(2020)Seger, Avin, Pearson, Briers, \'O~h\'Eigeartaigh, and Bacon}]{SegerEpistemicSecurity}
Seger, E.; Avin, S.; Pearson, G.; Briers, M.; \'O~h\'Eigeartaigh, S.; and Bacon, H. 2020.
\newblock Tackling threats to informed decision-making in democratic societies: {P}romoting epistemic security in a technologicall-advanced world.
\newblock Technical report, The Alan Turing Institute, Defence and Security Programme.

\bibitem[{Sierra et~al.(2021)Sierra, Osman, Noriega, Sabater-Mir, and Perelló}]{sierravalue2021}
Sierra, C.; Osman, N.; Noriega, P.; Sabater-Mir, J.; and Perelló, A. 2021.
\newblock Value alignment: a formal approach.
\newblock ArXiv:2110.09240, arXiv:2110.09240.

\bibitem[{Vezhnevets et~al.(2023)Vezhnevets, Agapiou, Aharon, Ziv, Matyas, Du{\'e}{\~n}ez-Guzm{\'a}n, Cunningham, Osindero, Karmon, and Leibo}]{vezhnevets2023generative}
Vezhnevets, A.~S.; Agapiou, J.~P.; Aharon, A.; Ziv, R.; Matyas, J.; Du{\'e}{\~n}ez-Guzm{\'a}n, E.~A.; Cunningham, W.~A.; Osindero, S.; Karmon, D.; and Leibo, J.~Z. 2023.
\newblock Generative agent-based modeling with actions grounded in physical, social, or digital space using Concordia.
\newblock \emph{arXiv preprint arXiv:2312.03664}.

\end{thebibliography}

\appendix
\section{Assumptions and Restrictions}
In the paper as a whole, we assume:
\begin{enumerate} 
\item Agents can be either human or AI.
\item Conflict between goals is symmetric:
\begin{align*}
    c(g_x, g_y) = c(g_y, g_x)
\end{align*}
\item Conflict values range from 0 (completely compatible) to 1 (completely incompatible).
\item Each agent assigns a weight to each problem area, representing its importance to that agent.
\item Weights are non-negative and can sum to more than 1 across problem areas.
\item Each goal considered must be held by at least one agent; otherwise, it should be removed from $\mathcal{G}$. Thus, in each problem area, $|\mathcal{G}| \leq |\mathcal{A}|$.
\item The model assumes a static snapshot of agents' goals and doesn't account for dynamic changes over time. However, time-series misalignment data can be created by analyzing misalignment repeatedly over time.
\item The model doesn't address potential synergies or positive interactions between different goals.
\item The model doesn't consider hierarchical relationships or power dynamics between agents.
\end{enumerate}

We also impose the following constraints:
\begin{enumerate}
\item Each agent holds no more than one goal per problem area.
\item A lack of a goal ($g_0$) is not in conflict with any explicit goal.
\item All agents are equally likely to be selected.
\end{enumerate}

These constraints mostly hold true in real life, but may have exceptions. For example, human individuals seem to hold contradictory goals in a single problem area. For example, a person may feel conflicted about eating a cookie, because they want to eat less sugar for dietary health reasons, but at the same time they might want to eat the cookie because they would enjoy the experience of eating it. 

We can fit this into our model more easily by splitting the problem area of "eating a cookie" into two -- "consuming sugar" and "having the experience of eating a cookie". We hypothesize that for any problem area, if it is possible for one person to hold conflicting preferences as to the resolution of that problem, it is possible to split the problem into two non-conflicting problems, making the situation once again compatible with our model. To address the original problem -- whether to eat a cookie, in this case -- we propose that a final "judgment" could be determined by aggregating the sub-problem preferences and scaling according to their respective weights. We leave the testing of our problem-splitting hypothesis and of related aggregation methods for future work.

As for the second constraint (that a lack of a goal is not in conflict with any explicit goal), this might be false in cases where an agent's goal is related to persuasion, since that implies intending to induce a certain goal or goals in other agents. To apply our model to cases like this, simply allow conflict values between $g_0$ and other goals; this is compatible with our model in theory, but excluded from our experiments for simplicity since it is not relevant to most cases. For the case of persuasion in particular, it may be appropriate to add other meta-normative goals (i.e. people’s preferences w.r.t. which goals they hold). These could be goals like “Mind my own business,” “Intervene where there is opportunity to do good,” and “Intervene only if I know the people involved.” 

Regarding the third constraint, that all agents are equally likely to be selected, future work could remove this constraint by changing the agent-pair sampling component of the misalignment equation, or by artificially scaling up some parts of the population. However, since the practical value of our model is in aligning AI more fairly, we do not recommend this. 

We add an extra restriction in \S{3.5}: Mutually Exclusive Goals:
\begin{enumerate} 
\item Goals within a problem area are mutually exclusive.
\end{enumerate}

Although most real-world cases involve goals with less extreme conflict, the simplified calculation that emerges from this constraint is helpful for deriving the upper bounds of misalignment that emerge in uniformly distributed populations (as in \S{5.2}: Varying Goals, and  \S{5.4}: Goal Distribution).

\section{Theorems and Proofs/Derivations}

\subsection{Misalignment Probability}
\textbf{Theorem 1:} The probability of misalignment between two randomly selected agents $a_1$ and $a_2$ in a problem area $p$ is given by:
\begin{equation}
\begin{aligned}
    Pr(1 | \mathcal{A}, p) := & \,\, Pr(a_1, a_2 \text{ selected randomly from } \mathcal{A}, \\ 
    & a_1 \neq a_2) \cdot c(g^1_p, g^2_p)
\end{aligned}
\end{equation}
where $c(g^1_p, g^2_p)$ is the conflict between the goals of $a_1$ and $a_2$ in problem area $p$.

\textbf{Derivation:} 
The probability of misalignment is the product of two factors:
\begin{enumerate}
    \item The probability of selecting two different agents randomly from the population $\mathcal{A}$. %
    \item The probability of conflict between the goals of these two agents, given by $c(g^1_p, g^2_p)$.
\end{enumerate}
Multiplying these factors gives the probability that two randomly selected agents will have conflicting goals in problem area $p$.

\subsection{Misalignment with Mutually Exclusive Goals}
\textbf{Theorem 2:} For a problem area with mutually exclusive goals, the misalignment probability is:
\begin{equation} 
Pr(1|\mathcal{A},p) = \frac{\sum_{g \in \mathcal{G}}\sum_{g' \in \mathcal{G}, g' < g} 2|\mathcal{A}_g||\mathcal{A}_{g'}|}{|\mathcal{A}|(|\mathcal{A}|-1)}
\end{equation}
where $\mathcal{A}_g$ is the set of agents holding goal $g$.

\textbf{Derivation:} 
For mutually exclusive goals, we consider all pairs of distinct goals:
\begin{enumerate}
    \item The double summation $\sum_{g \in \mathcal{G}}\sum_{g' \in \mathcal{G}, g' < g}$ iterates over all unique pairs of goals.
    \item For each pair, $|\mathcal{A}_g||\mathcal{A}_{g'}|$ is the number of agent pairs with these goals.
    \item We multiply by 2 to account for order (agent 1 having goal $g$ and agent 2 having goal $g'$, and vice versa).
    \item Dividing by $|\mathcal{A}|(|\mathcal{A}|-1)$ normalizes by the total number of possible agent pairs.
\end{enumerate}
This gives the probability of selecting two agents with different goals, which equals the misalignment probability for mutually exclusive goals.

\subsection{Misalignment Upper Bound}
\textbf{Theorem 3:} The maximum misalignment for a uniformly distributed population with $|\hat{\mathcal{G}}|$ non-zero goals is:
\begin{equation}
\text{Max Misalignment} = \frac{|\mathcal{A}|(|\hat{\mathcal{G}}|-1)}{|\hat{\mathcal{G}}|(|\mathcal{A}|-1)}
\end{equation}

\textbf{Proof:} 
In a uniformly distributed population:
\begin{enumerate}
    \item Each goal $g \in \hat{\mathcal{G}}$ is held by exactly $\frac{|\mathcal{A}|}{|\hat{\mathcal{G}}|}$ agents
    \item For any agent $a \in \mathcal{A}$:
\begin{enumerate}
    \item $a$ shares a goal with $\frac{|\mathcal{A}|}{|\hat{\mathcal{G}}|} - 1$ other agents
    \item $a$ is misaligned with $|\mathcal{A}| - \frac{|\mathcal{A}|}{|\hat{\mathcal{G}}|}$ agents
\end{enumerate}
\end{enumerate}

Therefore, the probability of misalignment between two randomly selected agents is:
\begin{equation}
Pr(1|\mathcal{A},p) = \frac{|\mathcal{A}| - \frac{|\mathcal{A}|}{|\hat{\mathcal{G}}|}}{|\mathcal{A}|-1}
\end{equation}
Simplifying step by step:
\begin{align*}
Pr(1|\mathcal{A},p) &= \frac{|\mathcal{A}| - \frac{|\mathcal{A}|}{|\hat{\mathcal{G}}|}}{|\mathcal{A}|-1} \\
&= \frac{\frac{|\mathcal{A}||\hat{\mathcal{G}}|}{|\hat{\mathcal{G}}|} - \frac{|\mathcal{A}|}{|\hat{\mathcal{G}}|}}{|\mathcal{A}|-1} \quad \text{(multiply first term by $\frac{|\hat{\mathcal{G}}|}{|\hat{\mathcal{G}}|}$)} \\
&= \frac{\frac{|\mathcal{A}||\hat{\mathcal{G}}| - |\mathcal{A}|}{|\hat{\mathcal{G}}|}}{|\mathcal{A}|-1} \quad \text{(combine terms in numerator)} \\
&= \frac{|\mathcal{A}||\hat{\mathcal{G}}| - |\mathcal{A}|}{|\hat{\mathcal{G}}|(|\mathcal{A}|-1)} \quad \text{(multiply both parts by $|\hat{\mathcal{G}}|$)} \\
&= \frac{|\mathcal{A}|(|\hat{\mathcal{G}}|-1)}{|\hat{\mathcal{G}}|(|\mathcal{A}|-1)} \quad \text{(factor out $|\mathcal{A}|$ in numerator)}
\end{align*}
As Figure 4 shows, this probability represents the maximum possible misalignment because any non-uniform distribution would concentrate agents into fewer effective groups. This would increase the probability of randomly selecting aligned agents, thus reducing the overall misalignment probability. This also shows, 

\textbf{Theorem 4:} $\frac{|\hat{\mathcal{G}}|-1}{|\hat{\mathcal{G}}|}$ is the exclusive upper bound of misalignment for a population with $|\hat{\mathcal{G}}|$ uniformly distributed non-zero goals.

\textbf{Proof:} 
\begin{equation}
\lim_{|\mathcal{A}| \to \infty} \frac{|\mathcal{A}|(|\hat{\mathcal{G}}|-1)}{|\hat{\mathcal{G}}|(|\mathcal{A}|-1)} = \frac{|\hat{\mathcal{G}}|-1}{|\hat{\mathcal{G}}|}
\end{equation}

\subsection{Overall Misalignment Across Problem Areas}
\textbf{Theorem 5:} The overall misalignment across all problem areas $\mathcal{P}$ is:
\begin{equation}
\begin{aligned}
    Pr(1 | \mathcal{A}, \mathcal{P}) := \\
     \frac{1}{|\mathcal{P}|} \sum_{p \in \mathcal{P}} & \Big( Pr(a_1, a_2 \text{ selected randomly from } \mathcal{A} \\
    & a_1 \neq a_2) \cdot c(g^1_p, g^2_p) \cdot \sqrt{w^1_p \cdot w^2_p} \Big)
\end{aligned}
\end{equation}
where $w^i_p$ is the weight agent $i$ assigns to problem area $p \in \mathcal{P}$.

\textbf{Derivation}
We derive this formula by extending the misalignment calculation for a single problem area to multiple areas, incorporating weights to reflect the importance of each area to the agents involved.

\begin{enumerate}
    \item For each problem area $p$, we start with the basic misalignment probability from Theorem 1:
\begin{equation*}
\begin{aligned}
Pr(1 | \mathcal{A}, p) := & \,\, Pr(a_1, a_2 \text{ selected randomly from } \mathcal{A}, \\ 
& a_1 \neq a_2) \cdot c(g^1_p, g^2_p)
\end{aligned}
\end{equation*}

    \item We incorporate the importance of the problem area to both agents using their weights. We use the geometric mean of the weights:
    \[\sqrt{w^1_p \cdot w^2_p}\]
    
    We use the geometric mean because:
    \begin{itemize}
        \item It captures the "middle ground" between two agents' weightings.
        \item It reduces the impact on misalignment when one agent cares little about an area and another cares a lot.
        \item It's less sensitive to extreme values than the arithmetic mean.
        \item The square root keeps the result on the same scale as the original weights.
    \end{itemize}

    \item We multiply the basic misalignment probability by this geometric mean of weights:
    \[Pr(0 | \mathcal{A}, p) \cdot \sqrt{w^1_p \cdot w^2_p}\]

    \item To get the overall misalignment, we sum this weighted misalignment over all problem areas:
    \[\sum_{p \in \mathcal{P}} \Big(Pr(0 | \mathcal{A}, p) \cdot \sqrt{w^1_p \cdot w^2_p}\Big)\]

    \item Finally, we take the arithmetic mean across all problem areas by dividing by $|\mathcal{P}|$:
    \[\frac{1}{|\mathcal{P}|} \sum_{p \in \mathcal{P}} \Big(Pr(0 | \mathcal{A}, p) \cdot \sqrt{w^1_p \cdot w^2_p}\Big)\]
    
    We use the arithmetic mean because:
    \begin{itemize}
        \item It gives equal importance to each problem area in the overall score, such that the importance weights factored into the misalignment scores are what  determines the overall importance of each problem area.
        \item It allows for intuitive interpretation: equal increases and decreases in different areas cancel out.
        \item It's consistent with the idea that misalignment in each area contributes additively to overall misalignment.
    \end{itemize}
\end{enumerate}

Note that weights are allowed to sum to more than 1 across problem areas for each agent. This allows agents to care deeply about multiple areas without forcing trade-offs, reflecting real-world scenarios where agents can care strongly about many issues simultaneously.

\section{Applications of Misalignment Models}

Current state-of-the-art in alignment methods used to develop frontier AI models lack consistency, theoretical backing, and sociotechnical grounding. For example, attempts to align foundation models to a vague notion of “humanity” have in practice led to models that mainly align to Western culture and values \cite{johnson2022ghost}. By defining and measuring misalignment in this more complex and nuanced way, we present a strong theoretical basis to better measure and track the success (or lack thereof) of alignment methods in the real world; avoid reductionist traps that assume that humans are always aligned with each other; and verify that they work as intended.

Intuitively, research and policy that intends to “align AI” will be more successful if it is conducted with a clear understanding of what the term “align” means, what success vs. partial success vs. failure looks like, and how to measure the above.  Measuring alignment in an oversimplified manner may lead to mistaken assumptions that AI has been aligned where the reality is more complex -- for example, AI systems aligned to one country’s cultural preferences could be misaligned with another population that has different preferences. As shown in \S{6.1}: Shopping Recommender System, a misalignment analysis may reveal reward hacking scenarios as well as other concerns; our misalignment scores could potentially be used to improve system alignment by modifying certain aspects of the agentic setting.  Our hope is to steer researchers away from reductive and oversimplified notions of alignment/misalignment and towards a richer sociotechnical understanding of alignment. 

Other practical applications include designing new AI training methods for pluralistic alignment (e.g. applying multi-objective optimization methods to balance the preferences of multiple groups of annotators) and supplying policymakers with better levers for AI alignment governance (e.g. enabling the writing of policies that require new frontier AI models to achieve certain alignment scores on some set of problem areas and preference data).

\section{Hardware and Software}
Simulating our model is, computationally, very cheap. All experiments were run on a commercial laptop with the following hardware:
\begin{enumerate}
\item OS: Windows 10
\item CPU: AMD64 Family 25 Model 80 Stepping 0, AuthenticAMD
\item CPU Cores: 8
\item CPU Threads: 16
\item RAM: 15.40 GB
\item GPU: NVIDIA GeForce RTX 3080 Laptop GPU
\end{enumerate}

These are the software versions used for the experiments:
\begin{enumerate}
    \item Python Version: 3.12.4
    \item NumPy Version: 2.0.1
    \item Matplotlib Version: 3.9.1
\end{enumerate}

\end{document}